\documentstyle[12pt,a4wide]{article}
\newcommand{\nin}{\noindent}
\newcommand{\be}{\begin{equation}}
\newcommand{\ee}{\end{equation}}
\newcommand{\bea}{\begin{eqnarray}}
\newcommand{\eea}{\end{eqnarray}}
\newcommand{\br}{\hskip .25cm/\hskip -.25cm}

\newcommand{\nonu}{\nonumber\\}

\newcommand{\dg}{^\dagger}
\newcommand{\ol}{\overline} 

\begin{document} 

\begin{titlepage} 
\begin{flushright}
hep-th/0202047
\end{flushright} 
\vspace{0.1in} 
\begin{center} 
{\Large {\bf Composite Supersymmetries\\
 in low-dimensional systems}}
\end{center} 
\vspace{0.1in} 
\begin{center} 
{\bf Jean Alexandre}, {\bf Nick E. Mavromatos} and {\bf Sarben Sarkar} \\ 
\vspace{0.1in}
King's College London, Department of Physics - Theoretical Physics,
Strand, London WC2R 2LS, U.K. \\ 
\end{center} 

\vspace{0.2in}

{\small 
Starting from a $N=1$ scalar supermultiplet in 2+1 dimensions, 
we demonstrate explicitly
the appearance of induced
$N=1$ vector and scalar supermultiplets
of composite operators made out of 
the fundamental supersymmetric constituents. We discuss 
an extension 
to a  $N=2$ superalgebra with central extension, due to the existence
of topological currents in 2+1 dimensions.
As a specific model we consider a supersymmetric 
$CP^1$ $\sigma$-model as the constituent theory, 
and discuss the 
relevance of these results for
an effective description 
of the infrared dynamics
of planar high-temperature superconducting condensed matter models 
with quasiparticle excitations 
near
nodal points of their Fermi surface.}

\end{titlepage}

\section{Introduction}

Supersymmetry is a symmetry that links integer spin excitations (bosons) to
half-integer ones (fermions), and, in certain circumstances, provides a
satisfactory control of quantum fluctuations, to the extent that many exact
analytic results can be obtained on the phase structure of certain
relativistic field theories~\footnote{%
At present, non-relativistic theories with Galilean invariance are also
known to possess somew kind of supersymmetry, which however is not
sufficiently developed or undersdtood to allow exact results in the quantum
theory.}. From this point of view, supersymmetry is expected mainly to
appear, if at all, in theories of the fundamental interactions of Nature,
which are by construction relativistic. In fact supersymmetry was a helpful
tool for understanding the non-perturbative structure of strongly coupled
non-abelian gauge theories, of relevance to the weak and strong
interactions. Seiberg and Witten~\cite{seib} have managed very effectively
to exploit extended $N=2$ four-dimensional supersymmetry in $SU(2)$ gauge
theory so as to obtain complete non-perturbative information on the
phase-structure of the theory. This work opened the way to more exact
results, some of which pertained to theories in space-time dimensions
different from four. There has been considerable interest, for instance, in $%
N=2$ supersymmetric gauge theories in three-dimensional space-times~\cite%
{strassler}, where {\it some} exact non-perturbative information can also be
obtained. At present it is $N=1$ supersymmetry, which seems to be of
phenomenological relevance to Nature in four dimensions, and unfortunately
for it there are no such exact results in any dimension. Nevertheless, it
may be hoped that, by viewing $N=1$ supersymmetry as a broken phase of an
extended supersymmetry, some exact information on the phase structure can be
extracted. Attempts in this direction have already been made in the
literature, e.g. four-dimensional softly broken $N=2$ QCD~\cite{broken}.
Given the necessity of relativistic systems, dynamical supersymmetry was
thought to be irrelevant for theories of condensed matter. This is due to
the fact that the majority of systems in condensed matter involve
non-relativistic excitations around Fermi surfaces of finite-extent. For
high-temperature superconductors, however, which are known to be $d$-wave
superconductors (with nodes in the superconducting gap) it has been found
that as the doping is lowered and the compound becomes non-superconducting,
a pseudogap region is entered where there are {\it nodes }at the Fermi
surface. This prompted the use of relativistic theories for describing the
low-lying excitations near the nodes~\cite{dm,farakos,nodalliq}. As a
result, from the perspective of obtaining exact analytic information on the
phase structure of such systems it is interesting to examine situations
where the effective relativistic theories of such nodal excitations possess,
in addition, dynamical supersymmetry.The hope is that such results may also
be useful in yielding important information on excitations, even away from
the nodes. Such attempts have been initiated in \cite{dgm,sarkar} in the
context of the so-called spin-charge separation phase of planar
antiferromagnets~\cite{anderson}, where the fundamental degrees of freedom
are not elelctrons, but rather objects carrying only spin- (spinon) or only
electric-charge- (holon) degrees of freedom. In terms of such excitations
the continuum low-energy effective field theory describing the dynamics of
the antiferromagnets is a $CP^{1}$ $\sigma $-model coupled to Dirac
fermions. The $z$-magnons of the $\sigma $-model represent spinons, while
the Dirac fermions represent holon degrees of freedom, which carry electric
charge only~\cite{farakos}. The presence of $CP^{1}$ field theory
necessitates the appearance of non-propagating $U_{S}(1)$ gauge
interactions, which from the effective field theory viewpoint are viewed as
interactions with infinitely strong gauge coupling (due to the absence of
gauge kinetic (i.e. Maxwell) terms). In some cases~\cite{farakos}, the nodal
liquid is described in a `particle-hole' symmetric formalism which allows
the existence of {\it additional} non-Abelian gauge interactions of $SU(2)$
type, representing the redundancy inherent in the ansatz of the spinon and
holon degrees of freedom. In \cite{sarkar} it has been asserted that the
effective theory of holons can be represented by Dirac fermions near the
nodes of a Fermi surface provided the $SU(2)$ gauge interactions are weakly
coupled. This will be assumed throughout this work.

The infinitely strong $U_{S}(1)$ interactions can be integrated out to give~%
\cite{farakos} effective (dual) theories of spinon and holons comprising of 
{\it composite objects} made out of these fundamental constituents. These
represent the effective physical degrees of freedom of the nodal system. In  %
\cite{sarkar} the conditions for dynamical supersymmetry between spinon and
holon constituent degrees of freedom in such effective theories have been
analysed. The important point to note is that specific points in the
parameter space, depending on particular {\it doping concentrations}, allow
for a scalar $N=1$ three-dimensional supersymmetry between spinon ($z$%
-magnon) and holon ($\Psi $ Dirac fermions) degrees of freedom. \ In this we
study the induced {\it composite supersymmetry} in the effective theory. As
we shall show, the existence of a scalar $N=1$ supermultiplet for spinons
and holons suffices to induce $N=1$ scalar {\it and} vector supermultiplets
at a composite level in the dual theory (obtained from integration of the $%
U_{S}(1)$ interactions). It should be stressed that here we {\it do not}
supersymmetrize the $U_{S}(1)$ interactions. As we shall see, due to the
three dimensionality of the theory, it is possible obtain centrally extended 
$N=2$ supersymmetries at both the constituent and the composite level, with
the central extension being provided by appropriate topological `charges' to
be defined below~\cite{hlousek,edelstein}. The $N=2$ supersymmetry couples
the two $N=1$ multiplets of the composite supersymmetry via gauge
interactions, leading to interesting models. One possible model is the
abelian Higgs model, while under different circumstances, to be discussed in
detail later, a broken phase of a non-abelian $SU(2)$ Georgi-Glashow model
obtains. These two models can lead to different physics.

The structure of this article is as follows: in the next section we shall
discuss notations and conventions pertaining to a scalar $N=1$ supersymmetry
at a constituent spinon-holon level. In section 3 we shall review the
appearance of composite operators in these systems, discussed in \cite%
{farakos}. In section 4 we shall study an $N=1$ supersymmetry at a {\it %
composite} level, induced by the $N=1$ scalar supersymmetry at a {\it %
constituent }spinon-holon level~\cite{dgm,sarkar}, and in particular we
shall define the associated scalar supermultiplet (containing the scalar
composites of \cite{farakos}). In section 5 we shall discuss the induced $%
N=1 $ vector composite supermulitplet containing the vector composites of %
\cite{farakos}. In section 6 we shall discuss the elevation of these
supersymmetries in three dimensions into centrally extended ones with the
central extension being identified with the topological charge~\cite%
{hlousek,edelstein}. In section 7 we shall argue that the appropriate
effective composite theory is a $N=2$ supersymmetric abelian-Higgs-like
model, and discuss how this may lead to exact results on superconducting
properties of the model at the supersymmetric points. The latter implies the
existence of composite complex fermions, which are capable of ensuring the
exact masslessness of an appropriate unbroken abelian subgroup of $SU(2)$
which will be denoted by $U_{3}(1)$ and will be crucial for
superconductivity by a mechanism~promulgated before \cite{dm}. The
superconducting point of the nodal liquid is argued to be a {\it quantum
critical point} in the phase space of the model. A discussion on a pseudogap
phase will also be made. Conclusions and outlook will be presented in
section 8.

\section{Scalar $N=1$ Supersymmetry of spinon-holon constituents}

In \cite{sarkar} a detailed microscopic model
was presented, and the
associated continuum effective theory of doped antiferromagnets with
spin-charge separation was derived. 
The constituent spinon and holon degrees of freedom
formed a scalar $N=1$ {\it supermultiplet} at certain points of the
parameter space of the microscopic model, depending on the doping
concentration.

In this section we set up the notation and conventions in $2+1$ dimensions
that \ will be used in this paper and also review the earlier work. The
Dirac matrices are $2\times 2$ and are given by $\left( \gamma ^{0}\right)
_{\;\beta }^{\alpha }=-i\left(\sigma^{2}\right)_{\alpha \beta },\;\left( \gamma
^{1}\right) _{\;\beta }^{\alpha }=\left(\sigma^{1}\right)_{\alpha \beta }$ 
and $\left(\gamma ^{2}\right) _{\;\beta }^{\alpha }=
\left(\sigma^{3}\right)_{\alpha\beta}$ where $\sigma ^{a}$
are the Pauli matrices. Hence $\left\{ \gamma ^{\mu},\gamma ^{\nu}\right\}
=2\eta ^{\mu\nu}$ with $\eta ^{\mu\nu}=diag\left( -1,1,1\right) $ and 
$\mu,\nu=0,1,2$. We
raise and lower Dirac indices with the real antisymmetric tensor whose
components are given by $\epsilon ^{\alpha \beta }=(-\gamma ^{0})_{~\beta
}^{\alpha }=\epsilon _{\alpha \beta }$. This convention has the property
that real spinors remain real under the operation of raising and lowering
indices. 

It is important to notice that in the model of \cite{sarkar}, as a result of
the antiferromagnetic structure, there are {\it two}  complex
supermultiplets \footnote{%
Throughout this work we follow the representation of \cite{dm}, where
spinons are bosonic fields ($z$ magnons), while holons are electrically
charged fermions. For alternative approaches, where the holons are 
represented as bosons and the spinons as fermions see 
\cite{others}.}. 
Hence starting with 4 $\ N=1$ real scalar
supermultiplets, we can construct 2 $\ N=1$ complex scalar superfields $%
Z_{a} $, $a=1,2$, made out of complex supermultiplets $(z_{a},f_{a},\psi
_{a})$ which transform as

\begin{eqnarray}
\delta z_{a} &=&\overline{\varepsilon }\psi _{a}  \nonumber
\label{transfoinit} \\
\delta f_{a} &=&\overline{\varepsilon }\br{\partial}\psi _{a}  \nonumber \\
\delta \psi _{a} &=&\br{\partial}z_{a}\varepsilon +f_{a}\varepsilon ,
\end{eqnarray}

For a product of complex spinors it is
useful to note that
\bea\label{iden1}
\ol\psi_1\psi_2&=&\psi_1^{\star\alpha}\psi_{2\alpha}\nonu
&=&\psi_1^{\star\alpha}\psi_2^\beta\epsilon_{\beta\alpha}\nonu
&=&\psi_2^\beta\psi_1^{\star\alpha}\epsilon_{\alpha\beta}\nonu
&=&(\ol\psi_2\psi_1)^\star
\eea 

\nin and similarly that 

\be\label{iden2}
\ol\psi_1\gamma^\mu\psi_2=-(\ol\psi_2\gamma^\mu\psi_1)^\star
\ee 

\nin We also have the Fierz identity 

\bea\label{fiertz}
(\ol\chi_1\gamma_\mu\chi_2)(\ol\chi_3\gamma^\mu\chi_4)&=&
\left(\chi_1^{\star\alpha}(\gamma_\mu)^\beta_{~\alpha}\chi_{2\beta}\right)
\left(\chi_3^{\star\gamma}(\gamma^\mu)^\delta_{~\gamma}\chi_{4\delta}\right)\nonu
&=&-\chi_1^{\star\alpha}\chi_2^\beta\chi_{3\gamma}^\star\chi_{4\delta}
(\delta_\beta^\delta\delta_\alpha^\gamma+\delta_\beta^\gamma\delta_\alpha^\delta)\nonu
&=&-(\ol\chi_1\chi_4)(\ol\chi_3\chi_2)+(\ol\chi_1\chi_3^\star)(\ol\chi_4\chi_2^\star)^\star,
\eea 

\nin which is valid for any set of spinors $\chi_i$, $i=1,2,3,4$. We remind the
important property concerning the antisymmetric tensor $\epsilon_{\mu\nu\rho}$: 

\be\label{epsilon}
\epsilon_{\mu\nu\rho}\gamma^\rho=\left[\gamma_\nu,\gamma_\mu\right]=\gamma_{[\nu}\gamma_{\mu]}
\ee 

\nin Finally, if we write explicitely the $2\times 2$ matrix $\chi_1\ol\chi_2$, for any spinors $\chi_1$
and $\chi_2$, we find the following relations: 

\bea\label{propmatinit}
\chi_1\ol\chi_2+(\chi_2\ol\chi_1)^\star&=&-\left(\ol\chi_2\chi_1\right)\mathbf 1\nonu
\chi_1\ol\chi_2-(\chi_2\ol\chi_1)^\star&=&-\left(\ol\chi_2\gamma_\nu\chi_1\right)\gamma^\nu,
\eea 

\nin where {\bf 1} is the $2\times 2$ unit matrix in Dirac space. We thus find that for any
spinors $\chi_1$ and $\chi_2$ 

\be\label{propmat}
\chi_1\ol\chi_2=-\frac{1}{2}\left[\left(\ol\chi_2\chi_1\right)\mathbf{1}
+\left(\ol\chi_2\gamma_\nu\chi_1\right)\gamma^\nu\right]
\ee

The associated long wavelength dynamics is represented by a supersymmetric $%
\sigma $-model which is described by the superfield lagrangian: 
\begin{equation}
{\cal L}_{\sigma }=-\int d^{2}\theta \{D^{\alpha }Z_{a}^{\dagger }D_{\alpha
}Z_{a}+\left( Z_{a}^{\dagger }D^{\alpha }Z_{a}\right) \left( Z_{a}^{\dagger
}D_{\alpha }Z_{a}\right) \}  \label{smodellagr}
\end{equation}%
where $D_{\alpha }$ $\left( =\frac{\partial }{\partial \theta ^{\alpha }}%
+\theta ^{\beta }\br{\partial}_{\alpha \beta }\right) $ denotes the
supercovariant derivative. The superfield $Z$ satisfies the constraint $%
Z_{a}^{\dagger }Z_{a}=1$. In terms of components the constraint becomes: 
\begin{equation}
\sum_{a=1}^{2}z_{a}^{\dagger }z_{a}=1~,\quad \sum_{a=1}^{2}z_{a}^{\dagger
}\psi _{a}+h.c.=0~,\quad \sum_{a=1}^{2}\left( z_{a}^{\dagger
}f_{a}+f_{a}^{\dagger }z_{a}+\psi _{a}^{\dagger }\psi _{a}\right) =0.
\end{equation}

\section{Composite Meson Fields}

In the path integral the integration of the strongly-coupled $U_{S}(1)$
gauge interactions lead to the appearance of composite operators \cite%
{farakos}, of 'meson' and 'baryon' type, comprising of holon constituents.
In what follows we shall concentrate exclusively on {\it meson type}
operators~\cite{farakos}: if we organize the fermions (holons) into a
`colour' doublet $\chi =(\psi _{1},\psi _{2})$, built out of the two
(two-component) fermion species $\psi _{1}$ and $\psi _{2}$ (corresponding
to the two antiferromagnetic sublattices), then the bilinears ($a=1,2,3$)

\begin{eqnarray}  \label{triplets}
\Phi^a&=&\overline\chi \Sigma^a \chi  \nonumber \\
{\cal A}_\mu^a&=&\overline\chi i\Sigma^a_\mu \chi
\end{eqnarray}

\noindent transform as {\it triplets} under $SU(2)$, where $\Sigma
^{a}=\sigma ^{a}\otimes $ ${\bf 1}$ and $\Sigma _{\mu }^{a}=\sigma
^{a}\otimes \gamma _{\mu }$. On the other hand, the $SU(2)$ singlets are
given by the bilinears:

\begin{eqnarray}  \label{singlets}
{\cal S} &=&\overline\chi\chi  \nonumber \\
{\cal S}_\mu&=&\overline\chi i\Gamma_\mu\chi,
\end{eqnarray}

\noindent where $\Gamma _{\mu }=\gamma _{\mu }\otimes {\bf 1}$. Note that $%
{\cal S}$ is a parity violating mass term, and ${\cal S}_{\mu }$ a
four-component fermion-number current. On the other hand, the
parity-invariant fermion mass term transforms as a triplet under the group $%
SU(2)$. In models with a gauged $SU(2)$ symmetry among spinon and holons, as
required by a particle-hole symmetric spin-charge separation ansatz~\cite%
{farakos}, the dynamical generation of a parity-invariant holon mass-gap
will induce a dynamical breaking of the $SU(2)$ gauge symmetry down to a $%
U(1)$ compact subgroup. On ignoring {\it non-perturbative} effects, such a
phase would be superconducting by the anomaly mechanism of \cite{dm}.
However, due to the compact nature of the $U(1)\subset SU(2)$ subgroup that
is left unbroken, there are monopole-instanton effects that in general may
be responsible for a small but non-zero mass of the $U(1)$ gauge boson. This
will spoil superconductivity, thereby leading to a pseudogap phase for the
nodal liquid, as discussed in detail in \cite{farakos}. Superconductivity,
on the other hand, requires an exactly massless $U(1)$ photon. Such a case
arises in the Georgi-Glashow supersymmetric model of \cite{affleck}. In \cite%
{farakos} we did not discuss spinon contributions to the composite
operators, because we worked in a phase where the spinon gap was much larger
than the fermion (holon) dynamically generated mass gap. In the
supersymmetric situation \cite{sarkar}, any mass gaps of spinons and holons
are of equal magnitude. It is natural to consider the effect of spinon
composites could affect the mesons (\ref{triplets},\ref{singlets}). This can
be answered by invoking supersymmetry at a composite level. Since the mesons
are obtained in \cite{farakos} by integrating out a non-dynamical gauge
group, the natural thing to assume is that the supersymmetry at a
constituent magnon-holon level will be preserved at a composite level, and
this will {\it define} the appropriate spinon contributions to the
composites. In what follows we shall concentrate on the $SU(2)$ triplet
composites, which are the ones that could possibly couple to the gauge $%
SU(2) $ interactions present in the particle-hole symmetric formalism of
spin-charge separation \cite{farakos}. As we shall demonstrate below, the
bilinear composites can lead only to two {\it decoupled} $N=1$
supermultiplets at a composite level, a scalar and a vector. Coupling of
these two supermultiplets can only be seen if higher order constituent
operators appear in the definition of the composite operators (\ref{triplets}%
) and will be discussed in section 7. Such a coupling will allow the
appearance of a $N=2$ dynamical supersymmetry, with a central charge that
coincides with the topolgical charge~\cite{hlousek,edelstein} defined by the
vector fields in the three-dimensional case at hand.

\section{N=1 composite scalar supermultiplet} 

Let us now consider a scalar composite 
$\Phi =\overline{\psi }_{1}\psi _{2}$. 
$\Phi $ defined in this way is complex but has the generic composite
structure that we will study; the real scalar composites will be considered
at the end of this section. The supersymmetric transform of $\Phi $ induced
by the spinons and holons is $\delta \Phi =\overline{\varepsilon }\Psi $
where

\be
\Psi=(f_1^\star-\br\partial z_1^\star)\psi_2+(f_2-\br\partial z_2)\psi_1^\star.
\ee 

\nin The supersymmetric transform of $\Psi$ is $\delta\Psi=\tilde F\varepsilon+M^{(+)}\varepsilon$ where 

\bea
\tilde F&=&2(f_1^\star f_2-\partial_\mu z_1^\star\partial^\mu z_2)\nonu
M^{(+)}&=&-\left\{\gamma^\mu,\psi_2\partial_\mu\ol\psi_1\right\}
-\left\{\gamma^\mu,\psi_1\partial_\mu\ol\psi_2\right\}^\star.
\eea 

\nin Using Eq.(\ref{propmat}), we can also write 

\bea
M^{(+)}&=&\frac{1}{2}\left\{\gamma^\mu,\partial_\mu\ol\psi_1\psi_2\mathbf{1}
+\partial_\mu\ol\psi_1\gamma_\nu\psi_2\gamma^\nu\right\}\nonu
&&+\frac{1}{2}\left\{\gamma^\mu,\partial_\mu\ol\psi_2\psi_1\mathbf{1}
+\partial_\mu\ol\psi_2\gamma_\nu\psi_1\gamma^\nu\right\}^\star\nonu
&=&\gamma^\mu\partial_\mu\left(\ol\psi_1\psi_2\right)+\partial_\mu\ol\psi_1\gamma^\mu\psi_2
+\left(\partial_\mu\ol\psi_2\gamma^\mu\psi_1\right)^\star\nonu
&=&\br\partial\Phi-\ol\psi_1\br\partial\psi_2-\left(\ol\psi_2\br\partial\psi_1\right)^\star,
\eea 

\nin such that $\delta\Psi=F\varepsilon+\br\partial\Phi\varepsilon$ where 

\be
F=\tilde F-\ol\psi_1\br\partial\psi_2-\left(\ol\psi_2\br\partial\psi_1\right)^\star.
\ee 

\nin The supersymmetric transform of the auxilliary field $F$ is 

\bea
\delta F&=&2\ol\varepsilon\left[f_1^\star\br\partial\psi_2+f_2\br\partial\psi_1^\star
-\partial^\mu\psi_1^\star\partial_\mu z_2-\partial^\mu z_1^\star\partial_\mu\psi_2\right]\nonu
&&-\ol\varepsilon\left(f_1^\star-\br\partial z_1^\star\right)\br\partial\psi_2
-\ol\varepsilon\br\partial\left(\br\partial z_2-f_2\right)\psi_1^\star\nonu
&&-\ol\varepsilon\left(f_2-\br\partial z_2\right)\br\partial\psi_1^\star
-\ol\varepsilon\br\partial\left(\br\partial z_1^\star-f_1^\star\right)\psi_2\nonu
&=&\ol\varepsilon\gamma^\nu\left(f_1^\star-\br\partial z_1^\star\right)\partial_\nu\psi_2
+\ol\varepsilon\gamma^\nu\left(f_2-\br\partial z_2\right)\partial_\nu\psi_1^\star\nonu
&&+\ol\varepsilon\left(\br\partial f_1^\star-\Box z_1^\star\right)\psi_2
+\ol\varepsilon\left(\br\partial f_2-\Box z_2\right)\psi_1^\star\nonu
&=&\ol\varepsilon\br\partial\Psi,
\eea 

\nin which closes the superalgebra, since we have found the composite superpartners
$F$ and $\Psi$ of the field $\Phi$ such that 

\bea
\delta\Phi&=&\ol\varepsilon\Psi\nonu
\delta F&=&\ol\varepsilon\br\partial\Psi\nonu
\delta\Psi&=&\br\partial\Phi\varepsilon+F\varepsilon.
\eea 

\nin Such a closure of the scalar superalgebra is also valid for 
the $SU(2)$ real triplet $(a=1,2,3)$ 

\be
\Phi^a=\ol\chi\Sigma^a\chi,
\ee 

\nin and the $SU(2)$ real singlet

\be
{\cal S}=\ol\chi\chi,
\ee

\nin with the notations introduced in the previous section.

\section{N=1 composite vector supermultiplet} 

The supersymetric transformations for a $N=1$ Abelian vector 
supermultiplet are
in the Wess-Zumino gauge 

\bea\label{transfovec}
\delta a^\mu&=&\ol\varepsilon\gamma^\mu\lambda\nonu
\delta \lambda&=&-\frac{1}{2}f_{\mu\nu}\gamma^\mu\gamma^\nu\varepsilon,
\eea 

\nin where $\lambda$ is a real gaugino and 
$f^{\mu\nu}=\partial^{[\mu} a^{\nu]}$. 
Let us consider the composite vector field \be\label{compogauge}
A^\mu=\ol\psi_1\gamma^\mu\psi_2-z_1^\star\partial^\mu z_2+z_2\partial^\mu z_1^\star
\ee \nin and its {\it Abelian}
field strength $F^{\mu\nu}$. The vector field (\ref{compogauge}) is complex 
but
has the basic composite structure from which we will construct the real
gauge fields at the end of this section. We will look for the supersymmetric 
transform of $A^\mu$, but before that we will derive 
the following property of $F^{\mu\nu}$ (see Eq.(\ref{propF})):
using the Fiertz identity (\ref{fiertz}), we can see that for any real 
Grassmann
variables $\theta_1,\theta_2$ we have 

\bea
\ol\theta_1~\partial_\mu\left(\ol\psi_1\gamma_\nu\psi_2\right)
\gamma^{[\mu}\gamma^{\nu]}~\theta_2&=&
~~\left((\partial_\mu\ol\psi_1)\gamma_\nu\psi_2\right)
\left((\ol\theta_1\gamma^\mu)\gamma^\nu\theta_2\right)\\
&&+\left(\ol\psi_1\gamma_\nu(\partial_\mu\psi_2)\right)
\left((\ol\theta_1\gamma^\mu)\gamma^\nu\theta_2\right)\nonu
&&-\left((\partial_\nu\ol\psi_1)\gamma_\mu\psi_2\right)
\left(\ol\theta_1\gamma^\mu(\gamma^\nu\theta_2)\right)\nonu
&&-\left(\ol\psi_1\gamma_\mu(\partial_\nu\psi_2)\right)
\left(\ol\theta_1\gamma^\mu(\gamma^\nu\theta_2)\right)\nonu
&=&-\ol\theta_1\left[\gamma^\mu,\partial_\mu(\psi_2\ol\psi_1)
-\partial_\mu(\psi_1\ol\psi_2)^\star\right]\theta_2\nonumber,
\eea 

\nin and thus 

\be
\partial_\mu\left(\ol\psi_1\gamma_\nu\psi_2\right)\gamma^{[\mu}\gamma^{\nu]}=
-\partial_\mu\left[\gamma^\mu,\psi_2\ol\psi_1\right]
+\partial_\mu\left[\gamma^\mu,\psi_1\ol\psi_2\right]^\star.
\ee 

\nin We also have $\partial_\mu\left(z_a^\star\partial_\nu z_b\right)\gamma^{[\mu}\gamma^{\nu]}
=\left[\br\partial z_a^\star,\br\partial z_b\right]$, such that finally 

\be\label{propF}
F_{\mu\nu}\gamma^\mu\gamma^\nu=
-\partial_\mu\left[\gamma^\mu,\psi_2\ol\psi_1\right]
+\partial_\mu\left[\gamma^\mu,\psi_1\ol\psi_2\right]^\star
-2\left[\br\partial z_1^\star,\br\partial z_2\right].
\ee 

\nin The supersymmetric transform of $A^\mu$ is now 

\be\label{transcompvec}
\delta A^\mu=\ol\varepsilon\gamma^\mu\Lambda
-\ol\varepsilon~\partial^\mu\left(z_1^\star\psi_2-z_2\psi_1^\star\right),
\ee 

\nin where the gaugino $\Lambda$ is given by 

\be\label{defLambda}
\Lambda=(f_1^\star+\br\partial z_1^\star)\psi_2-(f_2+\br\partial z_2)\psi_1^\star,
\ee 

\nin and has the following supersymmetric transform 

\be\label{transLambda}
\delta\Lambda=M^{(-)}\varepsilon+\left[\br\partial z_1^\star,\br\partial z_2\right]\varepsilon,
\ee 

\nin with 

\be\label{Mmoins}
M^{(-)}=\left[\gamma^\mu,\psi_2\partial_\mu\ol\psi_1\right]
-\left[\gamma^\mu,\psi_1\partial_\mu\ol\psi_2\right]^\star.
\ee 

\nin Then, from the first of Eqs.(\ref{propmatinit}), we can also write 
that for any
spinor $\chi_1$ and $\chi_2$ 

\bea
\chi_1\ol\chi_2&=&-\left(\ol\chi_2\chi_1\right)\mathbf{1}
-(\chi_2\ol\chi_1)^\star\nonu
&=&\frac{1}{2}\left(\chi_1\ol\chi_2-\left(\ol\chi_2\chi_1\right)\mathbf{1}-(\chi_2\ol\chi_1)^\star\right)
\eea 

\nin such that 

\bea
M^{(-)}&=&\frac{1}{2}\left[\gamma^\mu,\psi_2\partial_\mu\ol\psi_1-(\partial_\mu\psi_1\ol\psi_2)^\star
-\left(\partial_\mu\ol\psi_1\psi_2\right)\mathbf{1}\right]\nonu
&&-\frac{1}{2}\left[\gamma^\mu,(\psi_1\partial_\mu\ol\psi_2)^\star-\partial_\mu\psi_2\ol\psi_1
-\left(\ol\psi_1\partial_\mu\psi_2\right)\mathbf{1}\right]\nonu
&=&\frac{1}{2}\left[\gamma^\mu,\partial_\mu(\psi_2\ol\psi_1)-\partial_\mu(\psi_1\ol\psi_2)^\star\right],
\eea 

\nin and we finally find with Eqs.(\ref{propF}) and (\ref{transLambda}) 

\be\label{transLambdafinale}
\delta\Lambda=-\frac{1}{2}F_{\mu\nu}\gamma^\mu\gamma^\nu\varepsilon.
\ee 

\nin Now let us come back to the transformation (\ref{transcompvec}). 
The latter is the expected one up to
an Abelian gauge transformation (total derivative): 
if we define the scalar $\rho$ such that its supersymmetric transform is
$\delta\rho=z_1^\star\ol\varepsilon\psi_2-z_2\ol\varepsilon\psi_1^\star$,
then we can write (\ref{transcompvec}) as 

\be
\delta\left(A^\mu+\partial^\mu\rho\right)=\varepsilon\gamma^\mu\Lambda,
\ee 

\nin such that together with (\ref{transLambdafinale}), we have
defined a $N=1$ vector composite supermultiplet, up to a gauge transformation.
This result can be applied to the $SU(2)$ real triplet $(a=1,2,3)$ 

\be
A_\mu^a=\ol\chi i\Sigma^a_\mu\chi-i\phi\dg\sigma^a\partial_\mu\phi+i\partial_\mu\phi\dg\sigma^a\phi,
\ee 

\nin and the $SU(2)$ real singlet

\be
B_\mu=\ol\chi i\Gamma_\mu\chi-i\phi\dg\partial_\mu\phi+i\partial_\mu\phi\dg\phi
\ee

\nin where $\phi=(z_1,z_2)$ is a two-component scalar field 
and the other notations were introduced in section 3. 
We remind that the field strength
$F_{\mu\nu}^a$ appearing then in Eq.(\ref{transLambdafinale}) 
is the {\it Abelian} one for each $a$
(i.e. we obtain three independent Abelian composite supermultiplets).

\section{Current supermultiplet and N=2 extended superalgebra} 

A specific feature of 2+1 dimensions is that we can always construct 
a topological conserved current $J_\mu$,
starting from a vector $A_\mu$. This topological current is 
\be
J_\mu=\epsilon_{\mu\rho\sigma}\partial^\rho A^\sigma.
\ee 
\nin In general it can be shown~\cite{hlousek} that
the current $J_\mu$ belongs to a new supermultiplet containing $A_\mu$ and
a spinorial current $\tilde S_\mu^\alpha$.
As a result, the $N=1$ supersymmetry is centrally extended, with the
central extension being provided by the topological charge associated
with the current $J_\mu$~\cite{hlousek}. Such a centrally extended supersymmetry characterizes the
constituent spin-charge separating
$CP^1$ supersymmetric model (\ref{smodellagr}) of \cite{sarkar},
according to the above mentioned general argument.
Details on this can be found in the litterature~\cite{hlousek}
and will not be repeated here. We only mention that, in 
this case, the
topological current belongs to a current supermultiplet
${\cal J}$, 
which is defined in terms of the $Z$-scalar-superfields incorporating
the magnon $z$ and spinon $\psi$ degrees of freedom:
\begin{equation}
{\cal J}_\alpha = \frac{1}{4\pi}
D^\beta D_\alpha Z^\dagger D_\beta Z~,
\label{topocurrent}
\end{equation}
where $\alpha,\beta$ are
spinorial indices, and $D$ denotes the chiral superspace derivative.
Due to the basic identity 
\begin{equation}
D^{\alpha }D^{\beta }D_{\alpha }=0
\end{equation}
one can see that this current
supermultiplet is identically conserved, 
i.e. $D^{\alpha }{\cal %
J}_{\alpha }=0$, 
without the use of the equations of
motion. This framework
can be applied at the composite level. There is again a current
supermultiplet involving the topological current constructed out of a
composite vector field $A_{\mu}^{a}$, $a=1,2,3$ in (\ref{triplets}). Although
there are three such currents, our interest is in the effective dual theory
of degrees of freedom which are massless at a perturbative level. As
discussed in detail in \cite{farakos} due to spontaneous symmetry-breaking,
only one of the vector fields $A_{\mu}^{a}$, $a=1,2,3$ will remain massless in
perturbation theory. From simulations this is also the case for the
composite $A_{\mu}$'s \cite{farakos}. Hence it is plausible to assume that the
vector composite fields are gauged. In what follows we shall denote this
vector field by the generic symbol $A_{\mu}$ without any component index. Our
analysis formally holds for any of the three components of the composite
vector field (\ref{triplets}). In particular, at a {\it composite} level we
are interested in demonstrating that the current 
$J_{\mu}=\epsilon_{\mu\nu\rho}\partial ^{\nu}A^{\rho}$ 
belongs to a new supermultiplet containing 
$A_{\mu}$
and a spinorial composite current $\tilde{S}_{\mu}^{\alpha }$. To show this
let us return to the supersymmetric transform of $A_{\mu}$ seen in the
previous section and define the spinorial current $\tilde{S}_{\mu}^{\alpha }$
by

\be\label{varA}
\delta A_\mu=\varepsilon^\alpha \tilde S_{\mu\alpha}=\ol\varepsilon \tilde S_\mu,
\ee 

\nin such that $\tilde S_\mu^\alpha=\gamma_\mu\Lambda^\alpha$.
Since the translations commute with the supersymetric transformations,
$\tilde S_\mu$ is actually a conserved current in the specific
gauge $\partial^\mu A_\mu=0$, in which by definition $A_\mu$ 
is also conserved and we 
have $\br\partial\Lambda=0$. 
From the discussion in the previous section it folows that  
the supersymmetric transform 
of $\tilde S_\mu^\alpha$ is

\bea\label{varStilde}
\delta \tilde S_\mu&=&-\frac{1}{2}\gamma_\mu F_{\rho\sigma}\gamma^\rho\gamma^\sigma\varepsilon\nonu
&=&-\frac{1}{2}\gamma_\mu\partial_\rho A_\sigma \gamma^{[\rho}\gamma^{\sigma]}\varepsilon\nonu
&=&\frac{1}{2}\gamma_\mu\partial_\rho A_\sigma \epsilon^{\rho\sigma\lambda}\gamma_\lambda\varepsilon\nonu
&=&\frac{1}{2}\gamma_\mu \br J\varepsilon.
\eea 

\nin Let us now look at the supersymmetric transform of $J_\mu$. 

\bea
\delta J_\mu&=&\epsilon_{\mu\rho\sigma}\partial^\rho\ol\varepsilon\gamma^\sigma\Lambda\nonu
&=&\ol\varepsilon\gamma_{[\rho}\gamma_{\mu]}\partial^\rho\Lambda\nonu
&=&\ol\varepsilon\br\partial S_\mu-\ol\varepsilon\gamma_\mu\br\partial \Lambda\nonu
&=&\ol\varepsilon\br\partial\tilde S_\mu
\eea 

\nin Thus in the gauge $\partial_\mu A^\mu=0$,
we have a conserved current supermultiplet satisfying the
supersymmetric transformations 

\bea\label{currentmultiplet}
\delta A_\mu&=&\ol\varepsilon \tilde S_\mu\nonu
\delta J_\mu&=&\ol\varepsilon\br\partial \tilde S_\mu\nonu
\delta \tilde S_\mu&=&\frac{1}{2}\gamma_\mu\br J\varepsilon.
\eea 

\nin Equations (\ref{currentmultiplet})
form a closed superalgebra,
since two successive supersymetric transformations applied on any of the
currents lead to a translation of this current: 

\bea\label{successive}
\delta_1\delta_2 A_\mu&=&\frac{1}{2}\ol\varepsilon_2\gamma_\mu\gamma_\nu\varepsilon_1
\epsilon^{\nu\rho\sigma}~\partial_\rho A_\sigma\nonu
\delta_1\delta_2 J_\mu&=&\frac{1}{2}\ol\varepsilon_2\gamma^\rho\gamma_\mu\gamma^\sigma\varepsilon_1
~\partial_\rho J_\sigma\nonu
\delta_1\delta_2 \tilde S_\mu&=&\frac{1}{2}\gamma_\mu\gamma^\nu\varepsilon_2
~\ol\varepsilon_1\br\partial \tilde S_\nu
\eea 

\nin It must be stressed that this result is independent
of the model that we consider for the composite fields, since 
the supersymmetric transformations
(\ref{transfoinit}) concern the original microscopic fields $z_a$ 
and $\psi_a$. 

Let us suppose that we have 
a Lagrangian for the composite vector field $A_\mu$.
Then we can define the corresponding
spinorial Noether's current $S_\mu^\alpha$ and the associated
supercharge $Q^\alpha=\int d^2x S_0^\alpha$. 
The new spinorial current $\tilde S_\mu^\alpha$
enables us to define a
new supercharge $\tilde Q^\alpha=\int d^2x \tilde S_0^\alpha$.
We will show that the anticommutator $\{Q^\alpha,\tilde Q^\beta\}$ contains
an antisymmetric part, proportionnal to the topological charge, which thus
defines a $N=2$ supersymmetric structure. {}From the infinitesimal 
transformation (\ref{varStilde}), 
we know that 

\be
\left\{\ol Q\varepsilon,\tilde S_\mu\right\}=\frac{1}{2}\gamma_\mu\br J\varepsilon,
\ee 

\nin and thus 

\be
\left\{ Q^\alpha,\tilde S^\beta_\mu\right\}\varepsilon^\gamma\epsilon_{\gamma\alpha}=
\frac{1}{2}(\gamma_\mu\br J)^\beta_{~\gamma}\varepsilon^\gamma.
\ee 

\nin Since the inverse of $\epsilon_{\gamma\alpha}$ is $\epsilon^{\alpha\gamma}=(-\gamma^0)^\alpha_{~\gamma}$,
we have then 

\be
\left\{ Q^\alpha,\tilde S^\beta_\mu\right\}=
-\frac{1}{2}(\gamma_\mu\br J)^\beta_{~\gamma}\epsilon^{\gamma\alpha}
\ee 

\nin The anticommutator between the two supergenerators is therefore 

\be\label{comutQQ}
\left\{ Q^\alpha,\tilde Q^\beta\right\}=\epsilon^{\alpha\beta}\frac{T}{2}
+(\gamma^k)^{\alpha\beta}\frac{1}{2}\int d^2x J_k.
\ee 

\nin where $k=1,2$ is a spatial index, and 
$T=\int d^2x J_0$ is the topological charge~\cite{hlousek, edelstein}. 
Since the matrices $\gamma^k$ are symmetric, (\ref{comutQQ}) also implies 

\be
\left\{ Q^\alpha,\tilde Q_\alpha\right\}=T,
\ee 

\nin which indicates a $N=2$ superalgebra structure. However, 
it must be stressed that
the mere appearance of a current supermultiplet
{\it does not} necessarily imply a trully $N=2$ extended dynamical
supersymmetry
in the physical spectrum of the model. This can only happen if the topological
current is an independent quantity. In the case of the composite
model discussed here, the topological current is constructed
out of the vector composites, and hence, in order to promote
the $N=2$ structures to a true dynamical supersymmetry of the spectrum
we must discuss in detail the dynamics of the composite
model, in terms of
the form of the associated Lagrangian. This will be the topic of the
next section, where we shall see that the existence of a true $N=2$
dynamical supersymmetry implies additional constraints for the coupling 
constants of the model~\cite{edelstein}.

\section{N=2 Supersymmetric Composite Abelian Higgs model}

As we have discussed in previous sections, the existence of constituent
supersymmetries implies, via the appropriate transformation laws, the
existence of {\it fermionic} composite excitations, which are parts of $N=1$
supersymmetric composite multiplets of scalar and vector type. At a bilinear
composite field level, to which we restricted our attention for the purposes
of the current work, we are unable to couple these two types of
supermultiplets. Such a coupling is provided by the gauge fields themselves.
As we shall argue below, the latter can be identified with the vector
composites. Indeed the viewing of the vector composites as gauge fields was
essential for the closure of the $N=1$ vector supermultiplets in section 5
since the closure of the vector supermultiplet under the constituent
supersymmetry transformations occurs only {\it up to abelian gauge
transformations}. Notice that the identification of the vector composites
with $SU(2)$ gauge fields leads to a dynamical breaking of the $SU(2)$ gauge
group to a compact abelian subgroup $U_{3}(1)$, in a way explained in detail
in \cite{farakos}, and reviewed below. Such a gauge coupling would
necessitate higher-than-bilinear constituent-field interactions among
spinons and holons in the definition of the various composite fields (\ref%
{triplets}), (\ref{singlets}). At a composite level this would imply six
order terms of constituent field operators in the constituent lagrangian,
which are irrelevant operators at low energies (i.e. in the infrared), at
least from a naive renormalization group point of view. One might then hope
that, as far as the underlying constituent theory of holons and spinons is
concerned, the infrared fixed point universality class will not be affected.

For instructive purposes we first review briefly the non-Abelian ansatz for
spin-charge separation of \cite{farakos}. The particle-hole symmetric ansatz
implies that the electron operators are composites of spinon and holon
constituents in the form: 
\begin{equation}
\xi \equiv \left( 
\begin{array}{cc}
c_{1}\qquad & c_{2} \\ 
c_{2}^{\dagger }\qquad & -c_{1}^{\dagger }%
\end{array}%
\right) =\left( 
\begin{array}{cc}
\chi \qquad & \chi _{2} \\ 
\chi _{2}^{\dagger }\qquad & -\chi _{1}^{\dagger }%
\end{array}%
\right) \left( 
\begin{array}{cc}
z_{1}\qquad & -{\overline{z}}_{2} \\ 
z_{2}\qquad & -{\overline{z}}_{1}%
\end{array}%
\right)  \label{ansatz}
\end{equation}%
which is valid at each lattice site of the underlying microscopic theory.
Here $c_{a},\,a=1,2$ are electron operators, $z_{a},\,a=1,2$ are magnons
(spinons), and $\chi _{a},\,a=1,2$ are Grassmann numbers on the lattice
representing electrically-charged holon excitations.

The dynamical nodal theory depends only on the operators $\xi $~\cite%
{farakos}, even away from the half-filling case, and thus there is a {\it %
dynamical} non-abelian gauge symmetry of $SU(2)$ type, expressed by the
invariance of the ansatz, and thus $\xi $, under simultaneous local $SU(2)$
rotations of the spinon and holon parts: 
\begin{eqnarray}
&~&\Psi _{i}\longrightarrow \Psi _{i}\;h_{i}\qquad {\rm where}\qquad \Psi
\equiv \left( 
\begin{array}{cc}
\chi _{1}\qquad  & \chi _{2} \\ 
\chi _{2}^{\dagger }\qquad  & -\chi _{1}^{\dagger }%
\end{array}%
\right)   \nonumber \\
&~&Z_{i}\longrightarrow h_{i}^{\dagger }\;Z_{i}\qquad {\rm where}\qquad
Z\equiv \left( 
\begin{array}{cc}
z_{1}\qquad  & -\overline{z}_{2} \\ 
z_{2}\qquad  & \overline{z}_{1}%
\end{array}%
\right) 
\end{eqnarray}%
where $h_{i}\in $ SU(2) and $i$ denotes lattice site.

In addition, there is a strongly-coupled dynamical $U_S(1)$ gauge symmetry
acting {\it only on} the $\psi$ fields, which is due to phase frustration
from holes moving in a spin background. Some arguments to justify this from
a microscopic point of view have been given in reference~\cite{dm,wong}.
Consequently, this symmetry is associated with exotic statistics of the
pertinent excitations~\cite{farakos}, which is an {\it exclusive feature of
the planar spatial geometry}. It should be noted that the existence of these
gauge symmetries together with the appropriate constraints on single
occupancy in antiferromagnetic materials, reduce the effective number of
degrees of freedom to the physical degrees of freedom of the system. The
ansatz (\ref{ansatz}) gives us a maximal symmetry $G=SU(2) \otimes U_S(1)$.

The dynamics will be specified by a hamiltonian which will determine any
spontaneous symmetry breaking that occurs. In the Hartree--Fock
approximation we obtain the Hamiltonian~\cite{farakos}: 
\begin{eqnarray}
&~&H_{{\rm HF}}=\sum_{\langle ij\rangle }{\rm Tr}\left[ \frac{8}{J}\Delta
_{ij}^{\dagger }\,\Delta _{ji}+\left( -t_{ij}(1+\sigma _{3})+\Delta
_{ij}\right) \Psi _{j}^{\dagger }\langle Z_{j}\,\overline{Z}_{i}\rangle \Psi
_{i}\right] +  \nonumber \\
&~&\sum_{\langle ij\rangle }{\rm Tr}\left[ \overline{Z}_{i}\langle \Psi
_{i}^{\dagger }\left( -t_{ij}(1+\sigma _{3})+\Delta _{ij}\right) \Psi
_{j}\rangle Z_{j}+{\it h.c.}\right] .  \label{hf}
\end{eqnarray}%
where $\Delta _{ij}$ is a Hubbard-Stratonovich field.

Using the gauge symmetries, then, of the (\ref{ansatz}) we can write 
\begin{equation}
\langle Z_{j}\overline{Z}_{i}\rangle \equiv |A_{1}|\,{\cal R}%
_{ij}\,U_{ij}~,~\langle \Psi _{i}^{\dagger }\left( -t_{ij}(1+\sigma
_{3})+\Delta _{ij}\right) \Psi _{j}\rangle \equiv |A_{2}|\,{\cal R}_{ij}.
\label{bilinears}
\end{equation}%
where ${\cal R}\in $ SU(2) and $U\in $ $U_{S}(1)$ are group elements~%
\footnote{%
The fact that apparently gauge non-invariant correlators are non-zero on the
lattice, is standard in gauge theories, and does not violate Elitzur's
theorem~\cite{elitzur}, due to the fact that in order to evaluate the
physical correlators one must follow a gauge-fixing procedure, which should
be done prior to any computation. The amplitudes $|A_{1}|$ and $|A_{2}|$ are
considered frozen which is a standard assumption in the gauge theory
approach to strongly correlated electron systems~\cite{anderson}.}. These
are phases of the above bilinears and, to a first (mean field) approximation
can be considered as {\it composites} of the constituent fields, such as
spinons $z$ and holons $\psi $. Fluctuations about such ground states can
then be considered by integrating over the constituent fields. Notice that
the fermionic Hartree-Fock bilinear in (\ref{bilinears}) is a singlet under
the abelian $U_{S}(1)$ group, as a result of the transformation of the holon
fields, $\Psi _{i}^{\dagger }\Psi _{j}\rightarrow U_{ij}\Psi _{i}^{\dagger
}\Psi _{j}$, as well as the transformations of $t_{ij}\rightarrow
U_{ji}^{\dagger }t_{ij}$ due to the spin frustration~\cite{dm,wong}, and of
the field $\Delta _{ij}\rightarrow U_{ji}^{\dagger }\Delta _{ij}$. This
implies that in the effective action (\ref{hf}), the $z$-magnons are
singlets under the $U_{S}(1)$ group, and couple only to the $SU(2)$ part~%
\cite{sarkardual}. The resulting continuum action of this part is then a $%
CP^{1}$ $\sigma $-model coupled to the full $SU(2)$ group. Details can be
found in \cite{sarkardual} and will not be repeated here.

As shown in \cite{farakos}, by integrating out the $U_{S}(1)$ factors, we
obtain an effective action which is a functional of 
\begin{equation}
y_{\left( i\right) \mu}=-\kappa ^{2}{\rm Tr}\left( M^{(i)}(-\gamma
_{\mu})R_{\left( i\right) \mu}M^{(i+{\hat{\mu}})}\gamma _{\mu}
R_{\left( i\right)\mu}^{\dagger }\right)   \label{yvariable}
\end{equation}%
where $i$ is a lattice site index of the microscopic theory, the constant $%
\kappa $ depends on the microscopic parameters of the underlying theory, and 
$M$ denotes a meson composite matrix which (in the bosonic theory of \cite%
{farakos}) can be expanded in terms of the various meson composites: 
\begin{equation}
M^{(i)}=\sum_{a=1}^{3}\Phi ^{a}(i)\sigma ^{a}+{\cal S}(i)1_{2}+
i\left( {\cal %
S}_{\mu}(i)\gamma ^{\mu}+\sum_{a=1}^{3}A_{\mu}^{a}\gamma ^{\mu}\sigma ^{a}\right) 
\end{equation}%
The notation for the various meson composites is as indicated in section 3.

For the $SU(2)$ group elements $R_{\left( i\right) l}$ we have the structure~%
\cite{farakos} 
\begin{equation}
R_{\left( i\right) \mu}={\rm cos}(|{\vec{a}}_{\left( i\right)\mu }|)+i{\vec{%
\sigma}}\cdot {\vec{a}}_{\left( i\right) \mu}{\rm sin}\left( |{\vec{a}}%
_{\left( i\right) \mu}|\right) /|{\vec{a}}_{\left( i\right) \mu}|
\end{equation}

Recalling that the gauge fields ${\vec{a}}_{\left( i\right) \mu}$ are actually
composites of the constituent holons and spinons, it is natural to {\it 
assume} that the vector $SU(2)$ gauge fields appearing in this way are {\it 
identical} to the corresponding vector meson fields $A_{\mu}^{a}$ 
(\ref{triplets}) obtained after integration of the strongly-coupled $U_{S}(1)$
abelian group~\cite{farakos}. In this way, integration over constituent
fields $z,\psi $ (after inclusion of appropriate Jacobian 
factors~\cite{farakos}) can be transformed  into an integration over 
meson fields, which
express quantum fluctuations about the appropriate ground state.

{}From the analysis of sections 4 and 5 it becomes clear that the {\it entire
composite multiplet} $M^{\left( i\right) }$ can be {\it supersymmetrized}.
As we shall argue below, in this way, the dynamics of the composite system
turns out to be equivalent to that of a $N=1$ supersymmetric Abelian-Higgs
model~\cite{edelstein}. The fact that the ground state of dynamical
supersymmetric systems yields zero energy by definition provides, then,
strong support for the physical correctness of the identification of the
local phase fluctuations of the Hartree-Fock ansatz with the vector
composites (\ref{triplets}).

Notice that, upon the identification of the gauge fields $a_{\mu}$ with the
vector composite fields $A_{\mu}^{a}$, $a=1,2,3$, we observe that there are
Maxwell kinetic terms for the $SU(2)$ gauge fields in the composite
effective theory, and in this way the $SU(2)$ interactions are promoted to
fully dynamical ones. Such terms come by considering $y_{\left( i\right)\mu}$
(\ref{yvariable}) for nearest neighbour lattice sites $i, i+{\hat \mu}$.

The Maxwell terms for the $SU(2)$ gauge interactions will appear in the form
of plaquette terms for the respective gauge link variables in the in the
lattice action: 
\begin{equation}
S_{{\rm SU(2)}}=\sum_{p}\left[ \beta _{{\rm 2}}(1-{\rm Tr}{\cal R}_{p})%
\right]   \label{plaquette}
\end{equation}%
where $p$ denotes plaquettes, the $\beta _{2}$ are inverse couplings, $\beta
_{2}\equiv \beta _{{\rm SU(2)}}\propto 1/g^{2}$. and ${\cal R}_{p}$, are a
product of the link variables over the plaquette $p$.

As argued in \cite{sarkar}, for weakly coupled $SU(2)$ interactions, the
Grassmann fermionic variables $\chi _{a}$ can be assembled into two Dirac
spinors 
\begin{equation}
\psi _{1}\equiv \left( 
\begin{array}{c}
\chi _{1} \\ 
\chi _{2}^{\dagger }%
\end{array}%
\right) ~,\qquad \psi _{2}\equiv \left( 
\begin{array}{c}
\chi _{2} \\ 
-\chi _{1}^{\dagger }%
\end{array}%
\right)   \label{spinors}
\end{equation}%
The inverse coupling $\beta _{2}$ must therefore be large on the other hand,
in order to have a Dirac spinor representation of the holons~\cite{sarkar}.
In our case this coupling is related to the constant $\kappa $ in (\ref%
{yvariable}). This is what will be assumed from now on.

The analysis of \cite{farakos} then, shows, that there is dynamical gauge
symmetry breaking in the composite lagrnagian in the phase where a
parity-invariant mass term appears, or equivalently in the phase where the
scalar composite field $\Phi _{3}$ (\ref{triplets}) acquires a non-trivial
vacuum expectation value (v.e.v.). $<\Phi _{3}>=u\neq 0$. Since $u$ is a
fermion (holon) condensate which is generated dynamically by means of the
strongly coupled $U_{S}(1)$ interactions, whose coupling is formally
infinite (of the order of the ultraviolet cut-off of the efective theory),
quantum fluctuations of the condensate $\Phi _{3}$ {\it will be completely
supressed}. Two of the SU(2) gauge bosons, $A_{\mu}^{1,2}$ acquire masses in
that case, proportional to the above vev, $\kappa ^{2}u^{2}\rightarrow
\infty $, and, hence, they will decouple from the effective theory of light
degrees of freedom, while the gauge field $A_{\mu}^{3}$ remains perturbatively
massless.

The symmetry breaking patterns of our theory, as well as the 
supersymmetry transformations derived in sections 4 and 5 
imply~\cite{farakos, sarkardual}
the following form of the effective composite action in the naive continuum 
limit: 
\begin{eqnarray} 
&~& I= \int d^3x \{ -\frac{1}{4}F_{\mu\nu}^2 
+ \frac{1}{2}(\partial_\mu {\cal S})^2 
+ \frac{1}{2}\left(D_\mu (A^3)\phi \right)^* \left(D^\mu (A^3)\phi \right)
+ \nonumber \\
&~& \frac{i}{2}{\overline \Lambda}\br \partial \Lambda + \frac{i}{2}{\overline 
\chi}\br \partial \chi + \frac{i}{2}{\overline \psi}\br D(A^3) \psi \}
\label{abelianhiggssusy} 
\end{eqnarray} 
where $\chi$ is the real superpartner of the scalar singlet ${\cal S}$,
and $\phi =\Phi_1 + i \Phi_2$ is a complex scalar, with $\psi $ its Dirac 
(complex) spinor superpartner, constructed out of the two real superpartners
of the scalar composites $\Phi_i, i=1,2$ of section 4. 
The Abelian gauge field $A^3_\mu$, whose field strength is denoted by 
$F_{\mu\nu}$,  is the unbroken subgroup of the original SU(2) gauge group. 

Notice that, despite the parity violating character of the 
composite excitations in the singlet supermultiplet, the corresponding 
terms in the action preserve parity~\footnote{For 
this reason we have not written explicitly the vector singlet
terms ${\cal S}_\mu$ and their superpartners. Due to their parity violating
nature, such terms should form a mutliplet by themselves, and
should not interact with the rest of the  terms. 
To understand this, 
one should notice that, as a result 
of the gauge nature of these excitations, if such interactions 
existed they 
should have the 
form: $[(\partial _\mu - {\cal S}_\mu)\phi]^2$, which 
would violate parity, and 
as such should not appear 
in our parity conserving composite effective action~\cite{farakos}.}. 

The action (\ref{abelianhiggssusy}) is invariant under the following 
$N=1$ supersymmetry:
\begin{eqnarray} 
&~& \delta \Lambda =-\frac{1}{2}\epsilon^{\mu\nu\rho}F_{\mu\nu}\gamma_\rho \varepsilon~, \qquad 
\delta A_\mu^3 ={\overline \varepsilon}\gamma_\mu \Lambda~, 
\nonumber \\
&~& \delta \psi = \gamma^\mu \varepsilon D_\mu(A^3) \phi~,
\qquad  \delta {\cal S}={\overline \varepsilon }\chi~, \qquad 
\delta \phi = {\overline \varepsilon }\psi 
\label{susytrnsf} 
\end{eqnarray} 
We observe that a major part of these transformations 
has already been derived in sections 4 and 5 
by demanding a scalar $N=1$ supersymmetry among the 
constituent spinon and holons. 

Unfortunately, however, at the level of bilinear composites, discussed in
section 3, one cannot see the gauge-potential dependent parts in the
corresponding gauge covariant derivatives in the transformation for $\delta
\psi$ in (\ref{susytrnsf}). The latter part involves four fermions and hence
it can only be derived from composite fields which contain higher order
products of constituent fields {\it in addition to} the quadratic bilinear
contributions studied here. As already mentioned, at a constituent level
such coupling terms, which couple the vector and scalar $N=1$
supermultiplets, appear to be irrelevant operators in a naive
renormalization group sense. We hope to come to a more systematic study of
such higher order terms in the composite operators in a future work.

Assuming for the purposes of this work that the $N=1$ supersymmetric action
of the composite fields has the form (\ref{abelianhiggssusy}), one may make
a comparison with the corresponding action appearing in \cite{edelstein}. In
this case we observe that the action does not have potential terms, in
contrast to \cite{edelstein}, which implies that it constitutes a specific
case of the actions of \cite{edelstein} corresponding to zero couplings $%
e=\lambda =0$ in the notation of that work, where $\lambda$ is a coupling
constant in front of a Higgs-type potential for the doublets $\phi$, and $e$
is a coupling for interacting terms of the form ${\overline \psi}\Lambda
\phi - {\overline \Lambda}\psi \phi^*$.

It is also important to notice that in our case with the trivial condition
on the couplings $e=\lambda =0$ the $N=1$ supersymmetry can be {\it elevated}
trivially to a $N=2$ supersymmetry in our case. This is because the two real
fermions $\Lambda $ and $\chi $ can be {\it assemblied} in a single but
complex Dirac fermion 
\begin{equation}
\Sigma \equiv \chi -i\Lambda 
\end{equation}%
with the corresponding kinetic term in the effective lagrnagian: 
\begin{equation}
\frac{i}{2}{\overline{\Sigma }}\br{\partial}\Sigma 
\end{equation}%
The action (\ref{abelianhiggssusy}) then has an $N=2$ supersymmetry
invariance, as shown in \cite{edelstein}. The corresponding infinitesimal
tranasformations are characterized by a complex parameter $\eta \equiv
\varepsilon e^{i\alpha }$, and they are equivalent to (\ref{susytrnsf}) with
real parameter $\varepsilon $, followed by a phase transformation for the
fermions $\Sigma \rightarrow e^{i\alpha }\Sigma $ and $\psi \rightarrow
e^{i\alpha }\psi $. In the more general case of \cite{edelstein}, where $%
e,\lambda \neq 0$ one has that the $N=2$ supersymmetry trnasformations imply
a {\it condition} on the couplings $e^{2}=8\lambda $ in the normalization of %
\cite{edelstein}.

In our case, the absence of potential terms is compatible with the fact that
the Abelian-Higgs model is obtained here from a spontaneous breakdown of the
Georgi-Glashow model. Indeed in that model $N=2$ supersymmetry implies a
vanishing superpotential~\cite{affleck}.

Notice that the analysis of \cite{edelstein}, shows that the Noether
supersymmetry currents corresponding to the model (\ref{abelianhiggssusy})
are such that the pertinent supercharges satisfy a $N=2$ algebra with
central charge given by the topological charge of the Abelian-Higgs model.
This is in accordance with the general arguments of \cite{hlousek},
discussed in some detail in section 6. For completeness we will explicitly
give the spinor charges%
\begin{equation}
Q=\int d^{2}x\left[ \left( -\frac{1}{2}\epsilon ^{\mu\nu\rho}F_{\mu\nu}
\gamma _{\rho}+i\br{\partial}{\cal S}\right) 
\gamma ^{0}\Sigma +i\left( \br{D}\left(
A^{3}\right) \phi \right) ^{\ast }\gamma ^{0}\psi \right] 
\end{equation}%
and correspondingly for $\overline{Q}$. The resulting centrally extended
superalgebra is \cite{edelstein}
\begin{equation}
\left\{ Q_{\alpha },\overline{Q}^{\beta }\right\} =2\left( \gamma
_{0}\right) _{\alpha }^{\;\beta }P^{0}+\delta _{\alpha }^{\;\beta }T
\end{equation}%
where $P^{0}=\int d^{2}x\left( \frac{1}{4}F_{ij}^{2}+\frac{1%
}{2}\left| D_{j}\left( A^{3}\right) \phi \right|^{2}\right) $,
with $i,j=1,2$, 
is the total energy and the central charge $T$ 
is the topological charge of the model, 
as discussed in the previous section.

We also notice that the existence of an $N=2$ dynamical supersymmetry at a
composite level is compatible with the elevation of the $N=1$ constituent
supersymmetry of the $CP^{1}$ $\sigma $-model to a $N=2$ supersymmetry, due
to the existence of topological currents~\cite{hlousek}, as explained in
section 6.

\section{Discussion: N=2 Supersymmetry implies Superconductivity}

The precise form of the effective theory is crucial for an understanding of
the phase structure of the nodal liquid. As already mentioned, at a
perturbative level, the gauge field $A^3_{\mu}$ is massless, and in fact the
theory is superconducting~\cite{farakos}. However this may not be in general
true when non-perturbative effects are taken into account, such as
monopoles, which are instantons in the (2+1)-dimensional theory~\cite%
{polyakov}. The monopole is a Euclidean configuration which behaves
asymptotically as 
\begin{equation}
\hat{\Phi}^{a}=\hat{r}^{a},\qquad \tilde{F}_{\mu}^{a}(x)=\frac{1}{g}
\frac{\hat{r}^{a}\hat{r}_{\mu}}{r^{2}}
\end{equation}
where the caret indicates a unit vector and the tilde indicates the dual
field tensor.

In the $N=2$ supersymmetric Georgi-Glashow model, or its $N=2$ Abelian
counterpart discussed here, the photon $A_{\mu}^{3}$ remains exactly massless,
even at a non-perturbative level, due to the presence of Dirac (complex)
fermions. This has been discussed in ~\cite{affleck}, and we shall not
repeat the discussion here. Thus the nodal liquid at the supersymmetric
point leads to the superconductivity mechanism proposed in \cite{dm}. It is
worthy of noticing that such a masslessness is a property of the existence
of complex fermions rather than composite supersymmetry. Simply, in our $N=2$
supersymemtric case the existence of complex conposite fermions is a
necessary consequence of the extended supersymmetry, and in this sense it is
the constituent supersymmetry of spinon and holons rather than the composite
one which guarantees superconductivity. Nevertheless, the existence of a $N=2
$ dynamical supersymmetric effective theory is important on its own in
yielding exact non perturbative results on the phase structure in the way
explained in~\cite{strassler}.

The important point is that such a supersymmetry-argument based mechanism
will work at strictly zero temperature, where supersymmetry is unbroken, and
hence the above considerations may point towards a {\it quantum critical
(superconducting) point} of the nodal liquid at the supersymmetric point of
the parameter space of the microscopic model~\cite{sarkar}~\footnote{%
On the other hand, in the case of the supersymmetric Abelian Higgs model
with {\it a non trivial Higgs potential for the scalars} of ~\cite{edelstein}%
, which is {\it not our case here}, there is a Higgs phase in which the
photon becomes massive. Such a phase would not be superconducting according
to the mechanism of \cite{dm}, since it would lead to a massive photon
exchange in the respective current-current correlator, and hence the Landau
criterion for superconductivity would be violated. But given the holon
mass-gap generation, such a phase would have been simply a pseudogap phase.}.

An interesting question concerns a transition from the superconducting to
the pseudogap phase of the nodal liquid in our case, where indeed the
supersymemtric point necessarily implies superconductivity of the nodal
liquid of excitations. This may be provided by a simple variation of the
doping concentration, which could take one away from the constituent
supersymmetric point. In that case there is an explicit breaking of
supersymmetry, and the masses of the spinon and holons are unequal. One may
then arrive at the situation of \cite{farakos}, where the spinons are very
massive and hence should be integrated out. In that case one is left with
composites made only of holons, the light degrees of freedom in the problem.
In such a case the effective theory is just the Georgi-Glashow model~\cite%
{sarkardual} without composite fermions (which may be thought of as being
``very massive and thus decoupled'' like the spinons). In such theories,
according to the standard analysis~\cite{polyakov}, non-perturbative effects
yield the photon $A_{\mu}^{3}$ a small mass, thereby leading to a pseudogap
phase. If the above scenario is true, it would imply that there is a quantum
critical point of the nodal liquid where the onset of superconductivity is
identified with the onset of centrally extended $N=2$ supersymmetries at
both constituent and composite levels.

Much more work is needed before one arrives at a complete understanding of
the underlying dynamics of the nodal liquids. In this work we have discussed
a possible r\^{o}le of supersymmetry in yielding some exact results
concerning the passage to a superconducting phase. In this respect we still
lack a complete {\it quantitiative} derivation of the extended
supersymmetric composite dynamics, given our present inabilitiy to handle
analytically higher order terms of constituent field operators in the
expressions for the compsoite fields. Such terms would ensure the extended $%
N=2$ supersymmetries arising from the gauge coupling of the $N=1$ composite
vector and scalar supermultiplets. Neverhteless, the exciting features on
the existence of extended supersymemtries, presented here, which could allow
some exact results on the phase structure of the nodal liquids to be
obtained, already open up interesting directions for theoretical research in
such systems. We hope to come back to such studies in the near future.

\section*{Acknowledgements}

The work is supported by the Leverhulme Trust (UK).

\end{document}